\documentclass{PoS}

\usepackage{amsfonts} % AMS
\usepackage{amssymb} % AMS
\usepackage{amsmath} % AMS
\usepackage{mathtools}
\usepackage{hhline}
\usepackage{graphicx}
\usepackage{placeins}

\newcommand{\bx}{{\bf x}}
\title{Expansion coefficient of the pseudo-scalar density using the gradient flow in lattice QCD}

\ShortTitle{Strange quark in nucleons using gradient flow}

\author{\speaker{J.G. Reyes}, J. Dragos, J. Kim, A. Shindler \\
        Facility for Rare Isotope Beams, Physics Department, Michigan State University, East Lansing, Michigan, USA\\
        E-mail: \email{reyes@frib.msu.edu}\\}

\author{T. Luu\\
  Institute for Advanced Simulation (IAS-4) FZJ, Germany \\
	}

\abstract{We use the Yang-Mills gradient flow to calculate the pseudo-scalar expansion coefficient $c_P^*(t_f)$.
  This quantity is a key ingredient to obtaining the chiral condensate and strange quark content of the nucleon using the Lattice QCD formulation,
  which can ultimately determine the spin independent (SI) elastic cross section of dark matter models
involving WIMP-nucleon interactions.
The goal, using the gradient flow, is to renormalize the chiral condensate and the strange content of the nucleon
without a power divergent subtraction.
Using Chiral symmetry and the small flow time expansion of the gradient flow,
the scalar density at zero flow time can be related to the pseudo-scalar density at non zero flow time.
By computing the flow time dependance of the pseudo-scalar density over multiple lattices box sizes,
lattice spacings and pion masses, we can obtain the scalar density of the nucleon.
Our lattice ensembles are $N_{f}=2+1$, PCAC-CS gauge field configurations, varying over $m_{\pi}\approx \{410,570,700\}$~MeV at
$a=0.0907$~fm, with additional ensembles that vary $a\approx \{0.1095,0.0936,0.0684\} $~fm at $m_{\pi} \approx 700$~MeV.
}

\FullConference{The 36th Annual International Symposium on Lattice Field Theory - LATTICE2018\\
		22-28 July, 2018\\
		Michigan State University, East Lansing, Michigan, USA.}

\begin{document}

\section{Introduction}

The study of nucleon structure and inter-nucleon interactions is important for applications in beyond the Standard Model physics (BSM).
The scalar density nucleon matrix element $\left\langle N| \bar{s} s | N \right\rangle$,
is of particular interest for dark matter searches \cite{Ellis:2008hf}. Dark matter is necessary to explain
multiple astrophysical phenomena, with one of the best candidates being weakly interacting massive particles (WIMPs)
theorized within the constrained minimal supersymmetric extension of the Standard Model (CMSSM)~\cite{Kurylov:2003ra,Savage:2015xta}.
Direct detection of such particles relies on the fact that they can interact with nuclei and that we can calculate such cross section accurately.

The scalar quark content of nucleons enters in the spin-independent part of WIMP-nucleon interactions with a Higgs boson as a mediator.
\begin{figure}
	\centering\includegraphics[width=5.5cm,height=5cm]{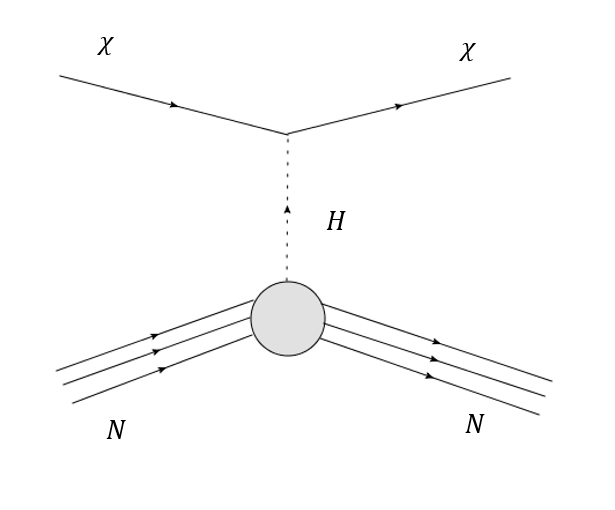}
	\caption{Wimp-nucleon interaction.}
	\label{fig1}
\end{figure}
The spin-independent scattering matrix elements are proportional
to the numbers of nucleons squared (i.e. coherent scattering) present in a nuclei.
We focus specifically on the strange quark contribution due to the higher mass relative to the valence quarks
but low enough mass as to be more prevalent in the quark sea (given that heavier quarks are strongly suppressed).
The dominant coupling of the Higgs particle to the nucleon will be accompanied by this scalar matrix element.
The spin independent (SI) elastic cross section from \cite{Bertone:2004pz} is
\begin{equation}
\sigma_{SI,\chi_N}\sim \bigg| \sum_{q_f} G_{q_f}({m^{2}_\chi}) \left\langle N| \bar{q}_f q_f | N \right\rangle \bigg|^{2},
\end{equation}
where $G_{q_f} $ is the effective coupling constant between the quark field $q_{f}$ and a WIMP, at scale $m_{\chi}$.
From this equation, we can see that small changes in the scalar content of the nucleon can lead to large changes in the cross section
 itself. In order to evaluate the impacts of dark matter detection experiments accurately, it is crucial to understand and minimize the hadronic uncertainties within the calculation. To interpret results from experiments, we have to model the WIMP-nucleus cross sections as accurately as possible.

 The strange quark content in nucleons has been computed in lattice QCD using a direct computational approach,
 and by using the Feynman-Hellman theorem ( see for example~\cite{Takeda:2010cw,Bali:2012rs,Freeman:2012ry,Dinter:2012tt,Liu:2017man}).
 Here we propose the use of the gradient flow as a tool to control the noise of these calculations
 and simplify the subtractions of infinities when going to the continuum due to the elimination
 (under the flow) of the mixing between operators with the identity.

\section{Pion Correlators and the Strange content of nucleons}

We apply the gradient flow to the quark and gauge fields as described in \cite{Luscher:2013cpa}.
For the fermion fields, the quark fields \(\psi(x)\) are transformed to a flowed quark field \(\chi(t_f,x)\), with \(\chi(0,x)=\psi(x)\).
The extra ``flow time'' parameter \(t_{f}\) denotes the amount of gradient flow applied.
In this proceedings we call the "flow-time radius" $\sqrt{8t_{f}}$ the root-mean-square radius of the flow evolution.
With this in mind, caution is needed when analyzing time scales \( \sqrt{8t_{f}}\leq t\leq T-\sqrt{8t_{f}}\).

We define the pseudo-scalar quark bilinear corresponding to the pion, using the flowed quark fields, as
\begin{equation}
  P^{ab}(t_f,x)=\bar{\chi}^a(t_f,x)\gamma_5\chi^b(t_f,x),
  \label{eq:P_tf}
\end{equation}
where \(a\) and \(b\) are the quark flavor index.
We calculate the two-point correlation function
\begin{equation}
  C(t_f,x_{0})= a^3 \sum_{\bx}\left\langle P^{ab}(t_f,x)P^{ba}(0,0)\right\rangle\,,
\end{equation}
and using the correspondent spectral decomposition we obtain
\begin{equation}
  C(t_f,x_{0}) = \sum_{m,n} \langle m |P^{ab}(t_{f},0) | n \rangle  \langle n | P^{ba}(0,0) | m \rangle e^{-x_0 M_n} e^{-(T-x_0 ) M_m},
\end{equation}
where \(T\) is the time extent of the lattice and $x=(x_0,\bx)$.
The lowest lying energy states for \(m,n\) which give a non-zero matrix element for \(P^{ab}(t_{f},0)\) are the vacuum $|0\rangle $
and a single pion state $|\pi\rangle $. Therefore, in the large Euclidean time approximation \(0 \ll x_{0} \ll T\), we have
\begin{align}\label{eq:twoptcorr}
  C(t_f,x_{0})=&\langle 0 | P^{ab}(t_{f},0) | \pi \rangle  \langle \pi | P^{ba}(0,0) | 0 \rangle  e^{-x_0 M_{\pi}} +
\langle \pi | P^{ab}(t_{f},0) | 0 \rangle  \langle 0 | P^{ba}(0,0) | \pi \rangle   e^{-(T-x_0 ) M_{\pi} } + \cdots \nonumber \\
=& 2G_{\pi,t_{f}}G_{\pi}e^{(T/2) M_{\pi}} \cosh[M_{\pi}(T/2-x_0)] + \cdots,
\end{align}
where we are neglecting excited states contributions and we assume that $\sqrt{8t_{f}}\ll x_{0} \ll T-\sqrt{8t_{f}}$.
The last condition guarantees that we can still treat the interpolating field~\eqref{eq:P_tf} local for the purposes of the spectral decomposition.
Results shown in sec.~\ref{sec:results} indicate that it is indeed the case alreasy for $x_0 \ge 2\sqrt{8t_f}$ and in same cases even for bigger values
of the flow-time radius.
The amplitudes are given by $G_{\pi}\equiv \langle  \pi |P^{ab}(0,0)| 0 \rangle $ ,
$G_{\pi,t_f} \equiv \langle 0 |P^{ba}(t_{f},0)| \pi\rangle $, and $M_{\pi}$ is the ground state mass of the pion.
As, under the assumptions discussed above, all the flow time dependence lies in the matrix element $G_{\pi,t_f}$,
we write the final form as
\begin{align}\label{eq:twoptcorr_fit}
C(t_f,x_{0})=& 2A(t_f) \cosh[M_{\pi}(T/2-x_0)],
\end{align}
where \(A(t_{f}) \equiv G_{\pi,t_f}G_{\pi}e^{(T/2) M_{\pi}}\).
The small flow time expansion~\cite{Luscher:2013vga} of the pseudo-scalar density
\begin{equation}
  P^{ab}(t_f,x)=c_P(t_f )P^{ab}(0,x)+O(t_f).
\end{equation}
implies that we can determine the expansion coefficient $c_P(t_f )$ from
\begin{equation}
  c_P(t_f )=\frac{Z^{2}_\chi A(t_f )}{Z_P A(0)}=\frac{Z^{2}_\chi G_{\pi,t_f}}{Z_P G_\pi} + O(t_f),
  \label{eq:c3}
\end{equation}
Where $Z_\chi$ is the wave function renormalization for flowed fermion field and $Z_P$
is the standard renormalization constant of $P^{ab}(0,x)$,
both determined in a given scheme.
%The renormalization of a composite field is $O_R(t_f)=(Z_\chi)^{\frac{1}{2}(n+\bar{n})}O(t_f)$
%where $n$ and $\bar{n}$ are the degrees of quark and antiquark fields respectively.
As described in ref.~\cite{Shindler:2014oha}, using the short flow-time expansion of the scalar density
together with chiral symmetry for the continuum theory, one obtains that the
vacuum subtracted correlation function
\begin{equation}
  C_{sub}(t_{f},x)=\left[ \left\langle N \bar{s}(t_f,x)s(t_f,x) \bar{N} \right\rangle -
    \left\langle \bar{s}(t_f,x)s(t_f,x)\right\rangle \left\langle N \bar{N} \right\rangle \right]\,,
\end{equation}
evaluated at non-vanishing flow-time is proportional to the physical scalar strange content, $C_{sub} (0,x)$, of the nucleon $N$
\begin{equation}
  C_{sub} (t_f,x)=c_P (t_f )  C_{sub} (0,x) + {\rm O}(t_f)
\end{equation}
up to corrections of higher powers in $t_f$ reflecting contributions from higher dimensional operators.

%Performing a small flow time expansion of the strange quark operators
%$O_s (t_f,x)=\bar{s}(t_f,x)s(t_f,x)$ and subtracting the irrelevant terms \cite{Shindler:2014oha}, we have
%\begin{equation}\label{eq:ft_exp}
%  O_s (t_f,x)=c_0 (t_f ) m_s+c_1 (t_f ) m_s (m_u^2+m_d^2+m_s^2 )+c_2 (t_f ) m_s^3+c_3 (t_f)O_s (0,x),
%\end{equation}
%for quark \(q\) masses \(m_{q}\), and coefficients \(c_{1,2,3}\) that run with the flow time.
%To avoid having to determine \(c_{1,2}\), we compute the expectation value of the difference of two quantities that differ only in
%the \(c_{3}\) term
%\begin{equation}\label{eq:sub_corr}
%  C_{sub}(t_{f},x)=\left[ \left\langle N O_s (t_f,x) N^{\dagger} \right\rangle -\left\langle O_s (t_f,x)\right\rangle \left\langle N N^{\dagger} \right\rangle \right].
%\end{equation}
%Applying \ref{eq:ft_exp} to \ref{eq:sub_corr}, we obtain a simple flow time dependence of
%\begin{equation}
%  C_{sub} (t,x)=c_3 (t_f )  C_{sub} (0,x)
%\end{equation}

Therefore we can determine the strange quark content between nucleons computing the subtracted matrix element $C_{sub}(t_{f},x)$
at non-vanishing flow time and then match it with the physical one at $t_f=0$ computing the expansion coefficient $c_P(t_f)$.
We also remark that one can perform the continuum limit at fixed physical value of the flow time $t_f$, without determining
the renormalization constant $Z_\chi$ in eq.~\eqref{eq:c3}.
In fact it simplifies with the renormalization constant of the scalar density at non-zero flow time.
Perturbation theory can give us a first estimate of the flow time dependence of $c_P(t_f)$~\cite{Luscher:2013cpa,Makino:2014taa}
\begin{equation}
  c_P(t_f)\sim (b_0 \bar{g}^{2})^{-8/9}\{1+O(\bar{g}^{2})\},
\end{equation}
where $\bar{g}$ is the renormalized coupling at a scale $\mu = 1/\sqrt{8t_f}$.

%We have $b_0$ being the one-loop coefficient of the QCD $\beta$ function and the coupling \cite{Hieda:2016lly} is
%\begin{equation}
%  \bar{g}^{2}\sim \frac{1}{b_0\log(1/(8t_f\Lambda^{2}))}.
%\end{equation}
% We also can consider computing the actual matrix elements
% \begin{equation}
%   \left\langle N O_s (t_f,x) N^{\dagger} \right\rangle ,
%   \qquad \left\langle O_s (t_f,x)\right\rangle \left\langle N N^{\dagger} \right\rangle,
% \end{equation}
% which would incorporate computing the \(<NO_{s}N^{\dagger}>\) from the (disconnected) strange quark contribution to the nucleon three-point
% correlation function, the strange quark condensate \(<O_{s}>\) and the nucleon two-point correlation function \(<NN^{\dagger}>\).
%But again, the gradient flow could be applied to the scalar strange quark bilinear \(O_{s}\) to renormalize the operator.

\section{Results}
\label{sec:results}
Determination of the coefficient $c_P(t_f)$ requires a precise calculation of the pion two-point correlation function, for which
the sink pion operator has the gradient flow applied to it.
These calculations were performed on 6 ensembles, three of which have varying spacings with almost equal volume
and pion mass \(m_{\pi}\).  These were used to study discretization effects. The remaining three have varying  \(m_{\pi}\)
and were used to perform a chiral extrapolation. Our simulation were performed with $N_{f}=2+1$ lattices from the ILDG \cite{Beckett:2009cb},
using a non-perturbatively \(O(a)\)-improved Wilson fermion action, on a Iwasaki gauge action.
A summary of the lattice parameters for all ensembles
 can be found in Table.~\ref{tab:ens_vals}.
\begin{table}
	\centering
	\begin{tabular}{|r ||c |c |c || c |c |c |}
		\hline
    Ensemble& $A_1$ & $A_2$ & $A_3$ & $M_1$ & $M_2$ & $M_3$ \\
		\hhline{|=||=|=|=||=|=|=|}
		$(L/a)^{3} \times T/a$ & $16^{3} \times 32$ &	$20^{3} \times 40$	& $28^{3} \times 56$ & $32^{3} \times 64$ &
    $32^{3} \times 64$	& $32^{3} \times 64$ \\
		$a$ [fm] & $0.1095$ & $0.0936$ & $0.0684$& $0.0907$ & $0.0907$ & $0.0907$ \\
		$L$ [fm] & $1.944$ & $1.960$ & $1.918$ & $2.9$ & $2.9$ & $2.9$ \\
		$2t_{cut}/T$ & $0.73$ & $0.62$ & $0.68$ & $0.72$ & $0.72$ & $0.72$ \\
		$N_{G}$ & $799$ & $799$ & $799$  & $399$ & $399$ & $399$ 		\\
		$M_\pi$ [MeV] & $738.1(6.6)$ & $674.3(2.8)$ & $659.0(2.7)$ & $699.7(3.6)$ & $574.6(3.3)$ & $409.1(2.5)$ \\
		$t_{0}/a^2$ &  $1.3629(16)$ & $2.2387(22)$ & $4.9886(65)$& $2.5377(16)$ & $2.4000(11)$ & $2.2591(12)$ \\
		\hline
	\end{tabular}
	\caption{Parameters for all ensembles used in this proceedings. The ensembles denoted as $A_{1,2,3}$ are used to study discretization effects and the
          ensembles denoted by $M_{1,2,3}$ the pion mass dependence of our results. $N_{G}$ refers to the number of
          gauge fields used, and the \(M_{\pi}\) values are obtained from the analysis of our 2-point function~\eqref{eq:twoptcorr} at $t_f=0$.
          The fit parameter $t_{cut}$ is described in the main text and it is compared in this table with half the time extent of the lattice.
  We also show the results for the scale parameter \(t_{0}/a^2\) for each ensemble (computed as described in \cite{Luscher:2010iy}).
          }
	\label{tab:ens_vals}
\end{table}

\begin{figure}
	\begin{minipage}{.5\textwidth}
		\centering
		\includegraphics[width = \textwidth]{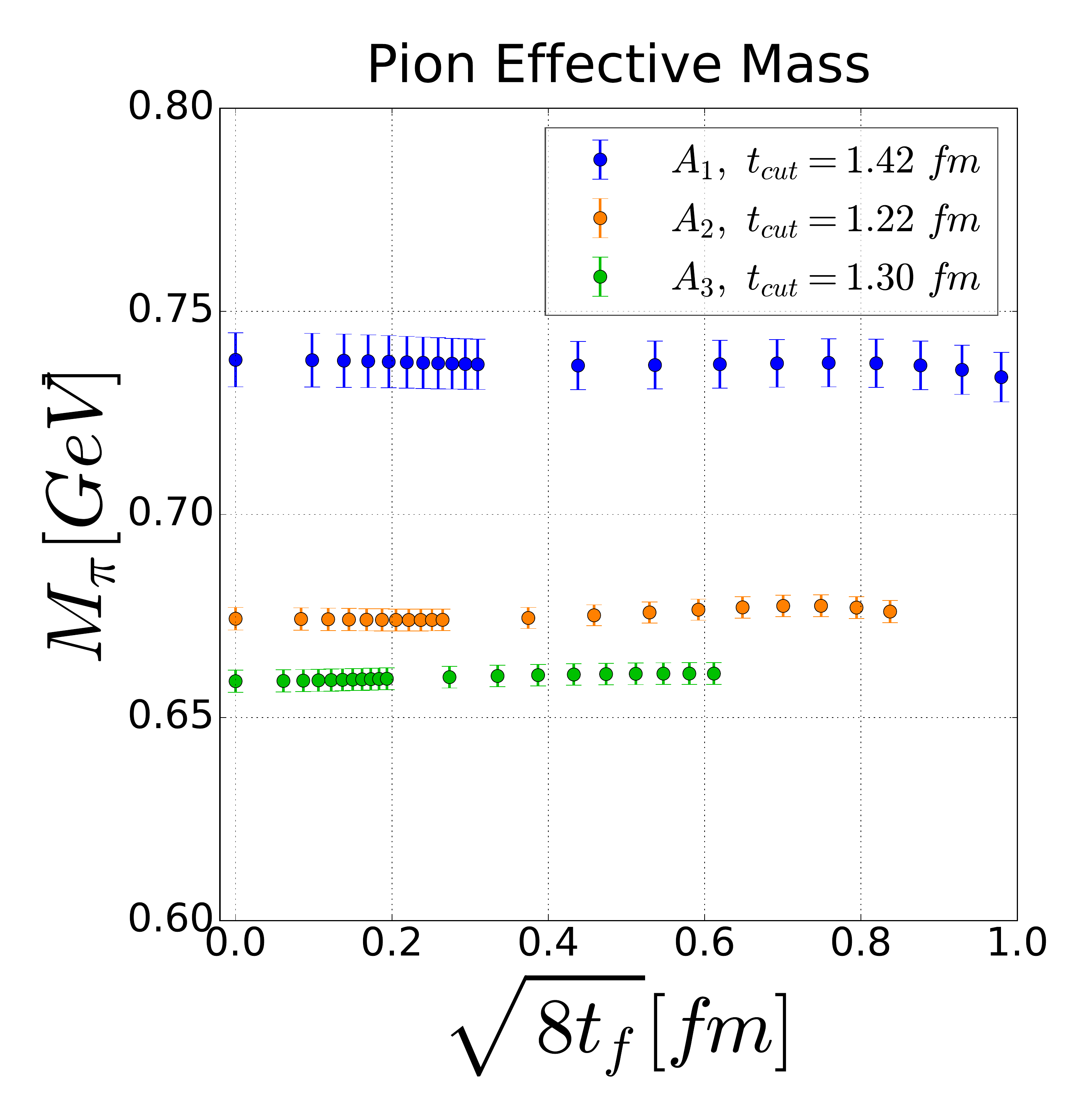}
	\end{minipage}%
	\begin{minipage}{.5\textwidth}
		\centering
		\includegraphics[width = \textwidth]{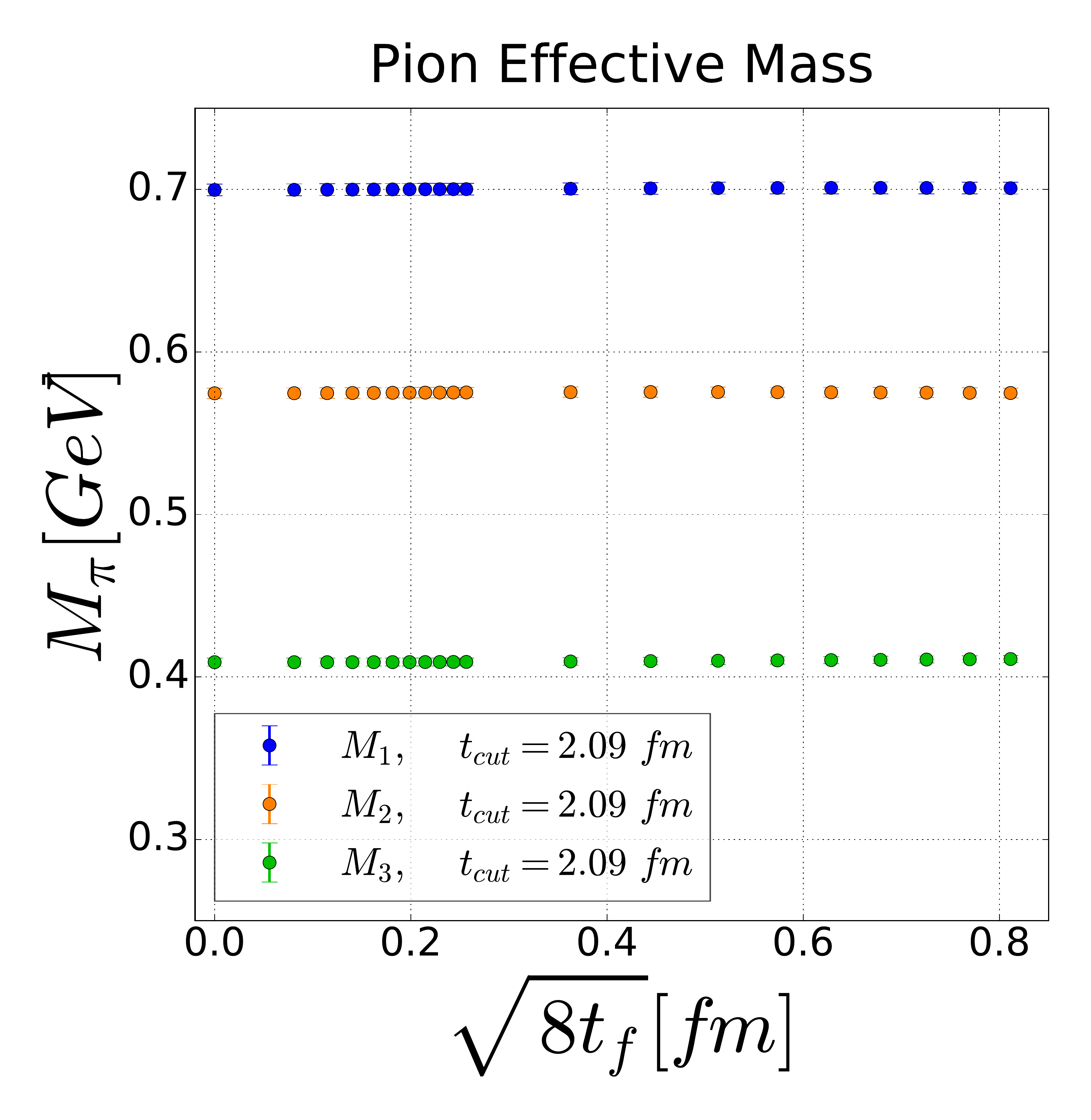}
	\end{minipage}%
		\label{fig:effmass}
	        \caption{Effective mass $M(t_{f})$ fit parameter results from fits over \(x_{0}\) in eq.~\ref{eq:effm}
                  for the $A_i$ (left) and $M_i$ (right) ensembles.
                  The value $t_{cut}$ refers to the region \(x_{0} \in [0,t_{cut}] \cup [T-t_{cut},T]\) \underline{excluded} from the fit.}
\end{figure}
\begin{figure}
	\begin{minipage}{.5\textwidth}
		\centering
		\includegraphics[width = \textwidth]{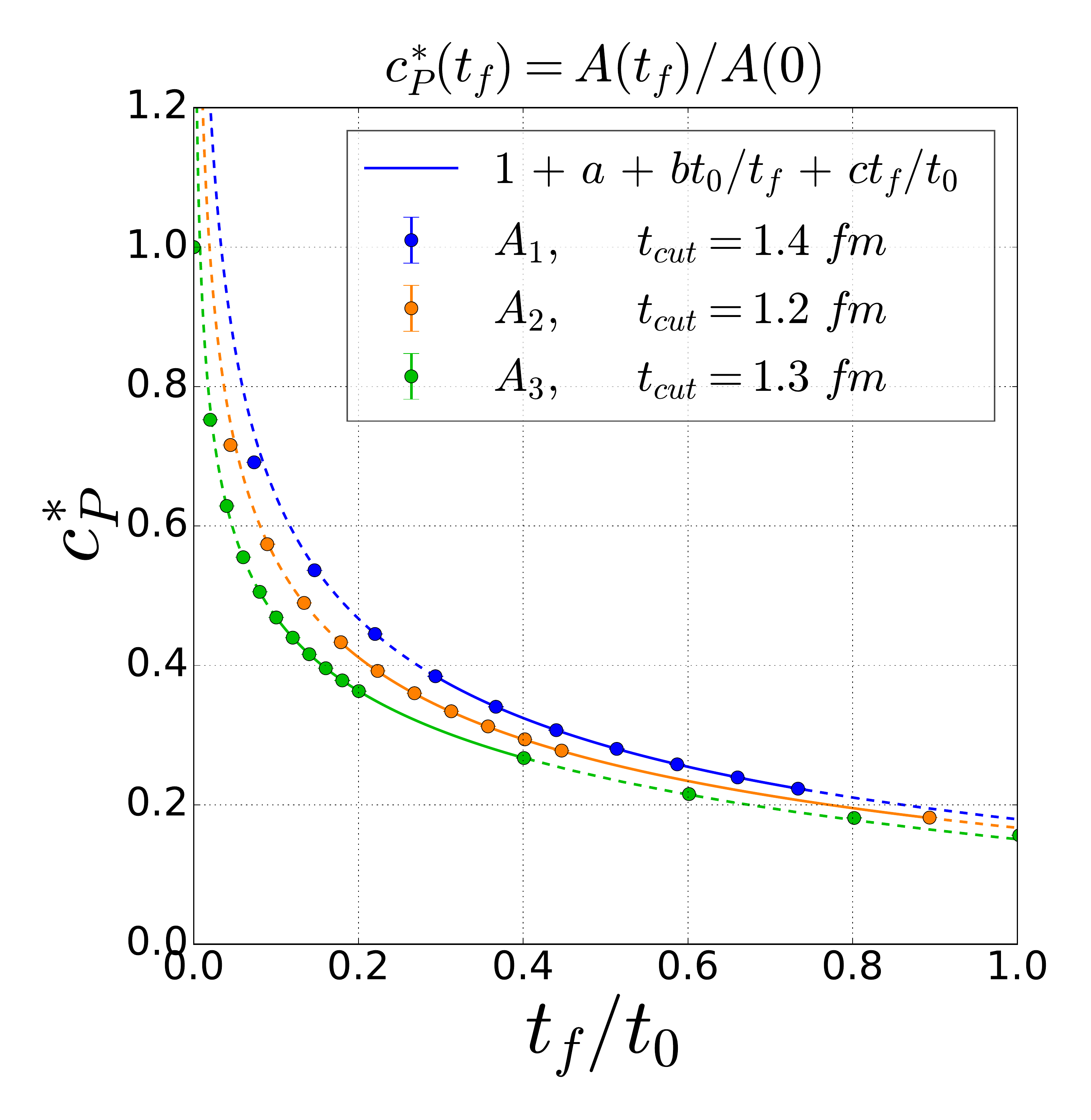}
	\end{minipage}%
	\begin{minipage}{.5\textwidth}
		\centering
		\includegraphics[width = \textwidth]{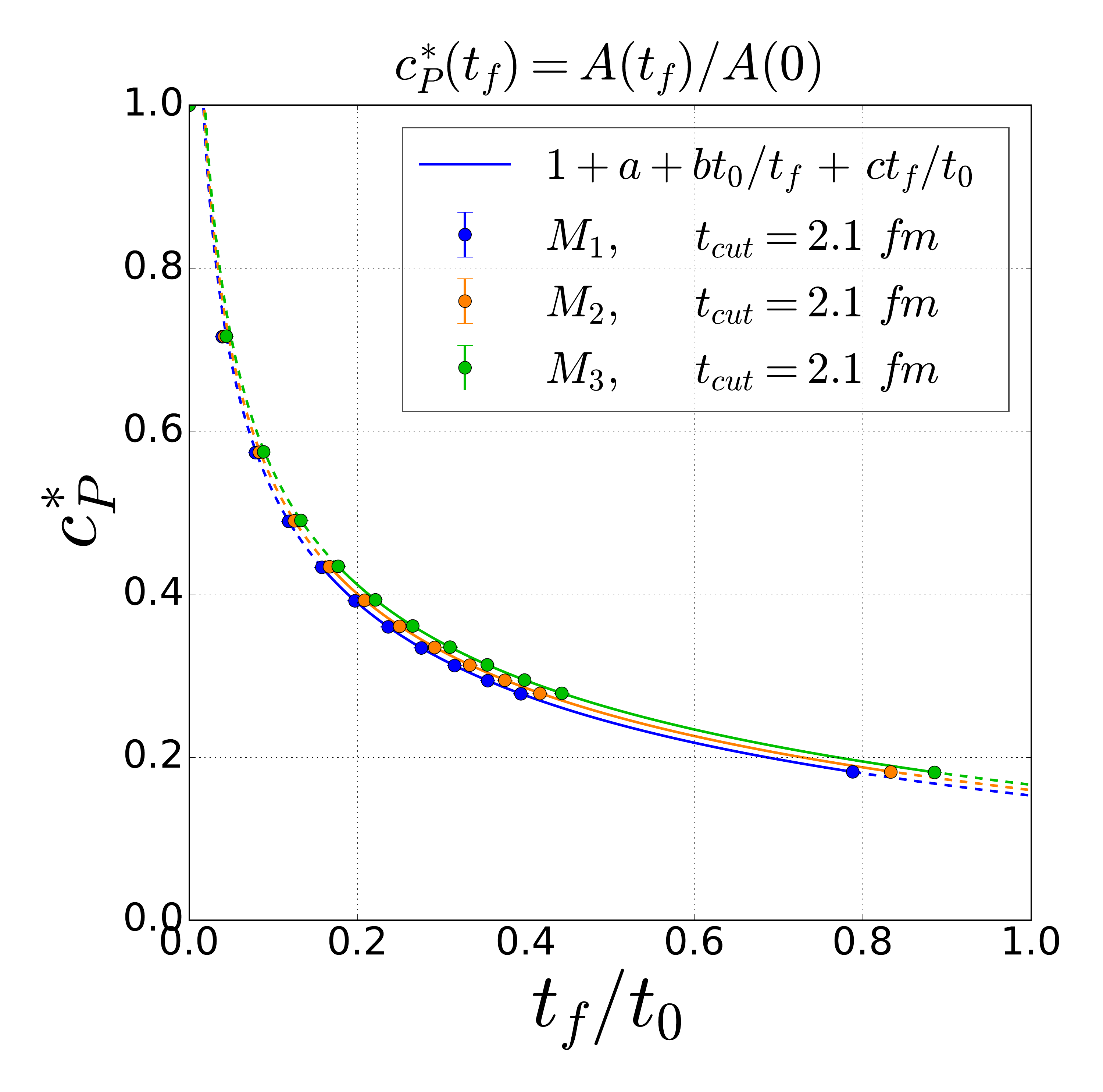}
	\end{minipage}%
	\caption{Ratio of flowed to un-flowed fit parameter \(A(t_{f})/A(0)\) for both the $A_i$ (left) and $M_i$ (right) ensembles.
          The legend value $t_{cut}$ refers to the region \(x_{0} \in [0,t_{cut}] \cup [T-t_{cut},T]\) \underline{excluded} from the fit.
          The solid lines indicate the fit ranges used to obtain the values in Tab.~\ref{tab:fit_res}.
          Dashed lines are shown just to visualize the agreement of the fit results with the data outside the fit range.}
		\label{fig:coeff}
\end{figure}

\begin{table}
	\centering
	\begin{tabular}{|r ||c |c |c |c ||}
		\hline
{} & $\chi^{2}_{pdf}$ &            $a$ &           $b$ &$c$\\
		\hhline{|=||=|=|=|=||}
$M_1$  &    $1.10(76)$ &    $-1.3284(16)$ &  $0.47895(56)$ &               $0.00227(98)$ \\
$M_2$  &    $1.5(1.1)$ &    $-1.3149(12)$ &  $0.48126(46)$ &  $-6.79(68) \times 10^{-3}$ \\
$M_3$  &    $0.56(52)$ &  $-1.30(0.00)$ &  $0.48489(29)$ &   $-1.61(4) \times 10^{-2}$ \\
% $M_1$  &    $1.10(76)$ &   $-1.3284(16)$ &  $0.038273(45)$ &    $0.028(12)$ \\
% $M_2$   &   $1.5(1.1)$ &    $-1.3149(12)$ &  $0.037925(36)$ &  $-0.0862(86)$ \\
% $M_3$   &  $0.56(52)$ &  $-1.30(00)$ &  $0.037637(22)$ &  $-0.2071(52)$ \\
		\hhline{|=||=|=|=|=||}
$A_1$     &  $1.47(29) \times 10^{-4}$ &              $-1.6202(23)$ &  $0.66870(90)$ &                 $0.1307(13)$ \\
$A_2$     &                $0.090(90)$ &              $-1.3283(16)$ &  $0.49440(59)$ &  $8.59(8.51) \times 10^{-4}$ \\
$A_3$     &                 $0.13(12)$ &  $-0.957(1)$ &  $0.30065(27)$ &    $-0.193(1)$ \\
% $A_1$ &  $1.47(29) \times 10^{-4}$ & $-1.6202(23)$ &  $0.050261(68)$ &    $1.739(17)$ \\
% $A_2$ &   $0.090(90)$ &   $-1.3283(16)$ &  $0.038895(46)$ &    $0.011(11)$ \\
% $A_3$ &   $0.13(12)$ &  $-0.957(1)$ &  $0.024703(22)$ &  $-2.3495(89)$ \\
% 		\hhline{|=||=|=|=|=||}
% cont. lim. & -  &  $-0.6958(22)$ &  $-0.48500(44)$ &  $-0.2987(13)$ \\
		\hline
	\end{tabular}
	\caption{Resulting fit parameters from fits of \(c^{*}_P(t_{f}) = 1+a+bt_{0}/t_{f} + ct_{f}/t_{0}\), shown in Fig.~\ref{fig:coeff}.
  % ``cont. lim.'' is a continuum limit of each fit parameter of the form
  % $f(m_{\pi}^{2},a^{2}/t_{0}) = f_{0} + f_{a}a^{2}/t_{0} + f_{m_{\pi}}m_{\pi}^{2}$.
}
	\label{tab:fit_res}
\end{table}
%
% \begin{table}
% 	\centering
% 	\begin{tabular}{|r ||c |c |c ||}
% 		\hline
%     & $a$ & $b$ & $c$ \\
% 		\hhline{|=||=|=|=||}
%     $M_1$ & $73.60(24)$ & $6.40(4) \times 10^{4}$ & $-855.9(3.1)$ \\
%     $M_2$ & $74.01(16)$ & $6.42(3) \times 10^{4}$ & $-854.1(2.3)$ \\
%     $M_3$ & $74.931(89)$ & $6.52(1) \times 10^{4}$ & $-855.9(1.2)$ \\
% 		\hhline{|=||=|=|=||}
%     $A_1$ & $88.31(11)$ & $9.19(2) \times 10^{4}$ & $-1027.1(1.1)$ \\
%     $A_2$ & $61.074(53)$ & $4.22(1) \times 10^{4}$ & $-678.13(83)$ \\
%     $A_3$ & $54.964(69)$ & $3.29(1) \times 10^{4}$ & $-585.29(78)$ \\
% 		% $a$ & $73.60(24)$ &	$74.01(16)$	& $74.931(89)$ & $88.31(11)$ & $61.074(53)$	& $54.964(69)$ \\
% 		% $b$ & $6.40(4) \times 10^{4}$ & $6.42(3) \times 10^{4}$ & $6.52(1) \times 10^{4}$& $9.19(2) \times 10^{4}$ &
%     % $4.22(1) \times 10^{4}$ & $3.29(1) \times 10^{4}$ \\
% 		% $c$ & $-855.9(3.1)$ & $-854.1(2.3)$ & $-855.9(1.2)$ & $-1027.1(1.1)$ & $-678.13(83)$ & $-585.29(78)$ \\
% 		\hline
% 	\end{tabular}
% 	\caption{Resulting fit parameters from fits of \(c^{*}(t_{f}) = 1 + a + \frac{b}{c+log(t_{f})}\), shown in Fig.~\ref{fig:coeff}.}
% 	\label{tab:fit_res}
% \end{table}

The first step after computing the two-point correlator with flowed sink defined in eq.~\eqref{eq:twoptcorr}, is to find a region
in source-sink separation \(x_{0}\) for which the condition \(0\ll x_{0}\ll T\) of ground-state dominance, and the flow time condition
\(\sqrt{8t_{f}}\ll x_{0} \ll T-\sqrt{8t_{f}}\) of field locality, are satisfied.
To be conservative we have taken the largest value of the flow time $t_f = \bar{t}_f$ we have used to compute the 2-point
function~\eqref{eq:twoptcorr} and we have determined the effective mass
\begin{equation}\label{eq:effm}
  M(t_{f}=\bar{t}_f,x_{0}) \ =\ \frac{1}{a}\log \left[ \frac{C(t_f=\bar{t}_f,x_{0})}{C(t_f=\bar{t}_f,x_{0}+a)}\right]\,.
\end{equation}
We find a region in $x_0$ that guarantees a ground-state dominance, for $t_f=\bar{t}_f$ and describe it with
the time exclusion parameter $t_{cut}$.
In other words, the region \(x_{0} \in [0,t_{cut}] \cup [T-t_{cut},T]\) is \underline{excluded}
from the fit in the two-point correlation function. We then keep fixed $t_{cut}$ and determine the
effective mass~\eqref{eq:effm} for smaller values of $t_f$.
The corresponding effective mass determination is shown in Fig.~\ref{fig:effmass}.
For fixed value of $t_{cut}$ we observe that the effective mass does not depend on the flow-time.
This indicates that we have found the window in $t_{cut} < x_0 < T - t_{cut}$ and in $ \sqrt{8 t_f} < x_0 < T - \sqrt{8 t_f}$,
where we can fit in $x_0$ the correlator~\ref{eq:twoptcorr_fit} with fit
parameters \(A(t_{f}),\ M_{\pi}\) for every flow time \(t_{f}\).

From the fit parameter \(A(t_{f})\) we construct the coefficient
\begin{equation}\label{eq:tf_exp}
  c^{*}_P(t_f) \equiv \frac{A(t_f)}{A(0)} = \frac{Z_P}{Z^{2}_\chi}c_P(t_f)\,.
\end{equation}
We present the results for $c^{*}_P(t_f)$ with respect to the flow time \(t_{f}/t_{0}\) in Fig.~\ref{fig:coeff}
for the $A_i$ (left) and $M_i$ (right) ensembles.
We introduce the fixed scale \(t_{0}\) which is computed using the gradient flow applied to the energy (using the clover definition for the gauge field tensor) for each ensemble. The method is described in detail in \cite{Luscher:2010iy}, and the individual values of \(t_{0}/a^2\) can be found in Table.~\ref{tab:ens_vals}.

From these results, we perform a phenomenological fit of the type
\({c^{*}_P(t_{f}) = 1+a+bt_{0}/t_{f} + ct_{f}/t_{0}}\)
% \(c^{*}(t_{f}) = 1 + a + \frac{b}{c+log(t_{f})}\)
in the flow-time ranges shown as solid lines in Fig.~\ref{fig:coeff} over \(t_{f}/t_{0}\).
The resulting fit parameters
\(a,\ b\) and \(c\)
% , as well at their continuum extrapolations,
 are shown in Table.~\ref{tab:fit_res}.
The first observation we make is that the continuum form of the flow-time dependence seems to be
largely affected by discretization effects. The coefficient $b$ in our fit formula should vanish in the continuum limit.
Any non-zero value of $b$ parametrizes a $1/t$ spurious dependence in the short flow-time expansion
of the pseudo-scalar density that should vanish in the continuum limit.
Additionally we observe cutoff effects in the coefficient $a$ which are presumably O($a^2 \Lambda^2$) unavoidable even
in the O($a$) improved theory.
We also observe substantial contributions from higher dimensional operators parametrized by the fit parameter $c$.
We observe a very small pion mass dependence in both fit parameters $a$ and $b$.

\FloatBarrier

\section{Conclusion}
In these proceedings we have presented the results of the pseudo-scalar expansion coefficient \(c_P^{*}(t_{f})\)
with respect to the gradient flow time \(t_{f}\) using ensembles with varying lattice spacing and pion mass.
We observe discretization effects for \(a<0.1095\) fm, which modify the expected form of the continuum short flow-time expansion.
It is unclear at the moment if those cutoff effects will be greatly reduced once the expansion coefficient is
combined with the bare subtracted matrix element evaluated at the same flow times.
We consider this as a first step for the determination of the scalar content of the nucleon without power divergent subtractions.

\bibliography{refs}
\end{document}